\documentclass[aps,prb,twocolumn,10pt,a4paper]{revtex4}
\usepackage{graphicx}
\usepackage{amsmath}
\usepackage{amssymb}
\usepackage{bm}

\begin{document}


\newcommand{\vt}{\tilde V}
\newcommand{\tk}{T_\mathrm{K}}
\newcommand{\boldk}{{\boldsymbol{\mathrm{k}}}}
\newcommand{\K}{\boldk}
\newcommand{\modmu}{|\mu|}
\newcommand{\hfmu}{|\mu_0|}
\newcommand{\bra}[1]{\langle #1 |}
\newcommand{\ket}[1]{| #1 \rangle}
\newcommand{\cre}[1]{c_{#1}^\dagger}
\newcommand{\ann}[1]{c_{#1}^{\phantom{\dagger}}}
\newcommand{\dbyd}[2]{\left(\frac{\partial #1}{\partial #2}\right)}
\newcommand{\sgn}{\operatorname{sgn}}
\newcommand{\E}{\mathrm{e}}
\newcommand{\D}{\;\mathrm{d}}
\newcommand{\I}{\mathrm{i}}
\newcommand{\ex}[1]{\langle #1 \rangle}
\newcommand{\tr}{\operatorname{Tr}}
\newcommand{\brapn}{\bra{\Psi_0^N}}
\newcommand{\ketpn}{\ket{\Psi_0^N}}
\newcommand{\deltar}{\Delta_\mathrm{R}}
\newcommand{\deltai}{\Delta_\mathrm{I}}
\newcommand{\deltao}{\Delta_0}
\newcommand{\sigr}{\Sigma^\mathrm{R}}
\newcommand{\sigi}{\Sigma^\mathrm{I}}
\newcommand{\gupa}{G_\uparrow^\mathrm{A}(\omega)}
\newcommand{\gupb}{G_\uparrow^\mathrm{B}(\omega)}
\newcommand{\gdna}{G_\downarrow^\mathrm{A}(\omega)}
\newcommand{\gdnb}{G_\downarrow^\mathrm{B}(\omega)}
\newcommand{\ga}[1]{G_{#1}^\mathrm{A}(\omega)}
\newcommand{\gb}[1]{G_{#1}^\mathrm{B}(\omega)}
\newcommand{\half}{\frac{1}{2}}
\newcommand{\niup}{\hat n_{\mathrm{i}\uparrow}}
\newcommand{\nidn}{\hat n_{\mathrm{i}\downarrow}}
\newcommand{\nisig}{\hat n_{\mathrm{i}\sigma}}
\newcommand{\scrgu}{\mathcal{G}_\uparrow}
\newcommand{\scrgd}{\mathcal{G}_\downarrow}
\newcommand{\ut}{\tilde U}
\newcommand{\dc}{\tilde \delta_\mathrm{c}}
\newcommand{\dt}{\tilde \delta}
\newcommand{\ot}{\tilde \omega}
\newcommand{\omegam}{\omega_\mathrm{m}}
\newcommand{\hc}{\hat{H}_\mathrm{c}}
\newcommand{\hh}{\hat{H}_\mathrm{h}}
\newcommand{\hp}{\hat{H}_\mathrm{p}}
\newcommand{\mub}{\mu_\mathrm{B}}
\newcommand{\epsi}{\epsilon_\I}
\newcommand{\epsit}{\tilde\epsi}
\newcommand{\simutoinf}{\overset{U\to\infty}{\sim}}
\newcommand{\simuttoinf}{\overset{\tilde U\to\infty}{\sim}}
\newcommand{\omo}{\ot^0_\mathrm{m}}
\newcommand{\om}{\omega_\mathrm{m}}
\newcommand{\eltup}{$| \epsilon | < U' $}
\newcommand{\egtup}{$| \epsilon | > U' $}
\newcommand{\eequp}{$| \epsilon | = U' $}
\newcommand{\eeqz}{$| \epsilon | = 0 $}
\newcommand{\eph}{$| \epsilon | = U' + \dfrac{1}{2} U $}
\newcommand{\gt}{\tilde\Gamma}
\newcommand{\jt}{\tilde J}
\newcommand{\jpt}{\tilde J'}
\newcommand{\ua}{\uparrow}
\newcommand{\da}{\downarrow}
\newcommand{\et}{\tilde \epsilon}
\newcommand{\gte}{\tilde \Gamma = \frac{\Gamma}{D} =}
\newcommand{\ute}{\tilde U = \frac{U}{\pi \Gamma} =}
\newcommand{\jte}{\tilde J = \frac{J}{D} =}
\newcommand{\jpte}{\tilde J' = \frac{J'}{D} =}
\newcommand{\phs}{\epsilon=-\frac{U}{2}}
\newcommand{\nimp}{\langle n_{\mathrm{d}} \rangle}
\newcommand{\thalf}{\tfrac{1}{2}}
\newcommand{\jpcte}{\tilde J'_{c}=\frac{J'_{c}}{D}=}
\newcommand{\ect}{\tilde \epsilon_{c}=\frac{\epsilon_{c}}{\pi \Gamma}=}
\newcommand{\nt}{\langle n_{2} \rangle}
\newcommand{\delt}{J-J'+\half \delta}
\newcommand{\delts}{\sqrt{(J-J'+\half \delta)^{2}+\frac{3}{4}\delta^{2}}}
\newcommand{\ttt}{\tilde t = \frac{t}{\pi \Gamma}}
\newcommand{\tpt}{\tilde t' = \frac{t'}{\pi \Gamma}}

\title{Two-channel Kondo physics in tunnel-coupled double quantum dots}


\author{Frederic W. Jayatilaka, Martin R. Galpin and David E. Logan}
\affiliation{Department of Chemistry, Physical and Theoretical Chemistry, Oxford University, South Parks Road, Oxford, OX1 3QZ, United Kingdom}


\date{\today}

\begin{abstract}
We investigate theoretically the possibility of observing two-channel Kondo (2CK) physics in tunnel-coupled double quantum dots (TCDQDs), at both zero and finite magnetic fields; taking the two-impurity Anderson model (2AIM) as the basic TCDQD model, together with effective low-energy models arising from it by 
Schrieffer-Wolff transformations to second and third order in the tunnel couplings. 
The models are studied primarily using Wilson's numerical renormalization group.
At zero-field our basic conclusion is that while 2CK physics arises in principle provided the system is sufficiently strongly-correlated, the temperature window over which it could be observed is much lower than is experimentally feasible. This finding disagrees with recent work on the problem, and we explain why. At finite field, we show that the quantum phase transition known to arise at zero-field in the two-impurity Kondo model (2IKM), with an essentially 2CK quantum critical point, persists at finite fields. This raises the prospect of access to 2CK physics by tuning a magnetic field, although preliminary investigation suggests this to be even less feasible than at zero field.
\end{abstract}


\maketitle
\section{Introduction}
Over the last decade or so, quantum dot devices\cite{kouw:97} have become increasingly important testbeds for the realization and controlled experimental study of correlated-electron phenomena. The spin-$\tfrac{1}{2}$ Kondo effect\cite{hews:93,kond:64} is the classic example, in which at low temperatures the spin degree of freedom of the dot is screened as a result of tunnel coupling to metallic leads. The rich physical behavior arising, in particular the strong many-body enhancement of the zero-bias conductance,\cite{gold:98,cron:98,vand:00} has stimulated the search for related phenomena in more complex device geometries: extensive work, both experimental and theoretical, has uncovered a wide range of examples, including orbital and $SU(4)$ Kondo effects,\cite{bord:03,choi:05,jari:05,maka:07,ande:08} underscreened Kondo behavior,\cite{roch:08,loga2:09,park:10,flor:11} and several Kondo effects induced by an applied magnetic field,\cite{pust:00,nyga:00,jari2:05,kiko:07,galp:10} to name but a few.

In this paper we consider a tunnel-coupled double quantum dot (TCDQD).\cite{vand:02} The system consists of two locally correlated and mutually tunnel-coupled quantum dots, positioned in series between
two metallic leads; and tunnel-coupled to them, such that 
current can flow through the system under a voltage bias applied to the leads. Experimentally, recent advances in nanofabrication have enabled construction of such systems in both carbon nanotube\cite{buit:08,inge:08, chur:09, sapm:06, grab:06} and semiconductor devices.\cite{vand:02, crai:04,jeon:01} 

From a theorist's perspective, the canonical model describing TCDQDs is the well known two-impurity Anderson model (2AIM).\cite{alex:64,gott:64,yama:79,jaya:82}
In a gate-voltage regime where each dot is effectively singly-occupied, the low-energy physics of the 2AIM is in turn embodied -- to leading order in the tunnel-couplings under a Schrieffer-Wolff transformation\cite{schr:66} -- in the 
two-impurity Kondo model (2IKM).\cite{jone:87,jone:88,jone:89,jone2:89,jone:91,affl:92,affl:95} The physics of the 2IKM is immensely rich.\cite{jone:87,jone:88,jone:89,jone2:89,jone:91,affl:92,affl:95} 
In particular, in the absence of an applied magnetic field,  it is well known  to contain a quantum phase transition, for which the quantum critical point is in essence a two-channel Kondo (2CK)  fixed point 
(FP).\cite{jone:88,jone:89,affl:92,affl:95,gan:95} That in turn raises the prospect of observing 2CK
physics in TCDQD systems, which recent theoretical work\cite{male:10} has suggested to be potentially viable.
This is a central 
issue considered in the present paper.

The simplest 
exemplar of 2CK physics is the 2CK model,\cite{nozi:80} consisting of a single spin-$\half$ coupled via antiferromagnetic Kondo exchange to two metallic leads, which compete to Kondo-screen the spin
and result in overscreening of it.\cite{nozi:80} In consequence, the 2CK ground state is a non-Fermi liquid, characterized by the stable infrared 2CK FP and exhibiting 
exotic physical properties such as a residual entropy of $\half\ln2$ ($k_{B} \equiv 1$).\cite{andr:84, affl:91}
The 2CK FP is however notoriously susceptible to destabilizing perturbations:\cite{pang:91, affl:92, sela:11} inter-lead charge transfer in particular -- 
as will always occur to some degree in a real device (and is inherently contained in the 2AIM) -- is well known to destabilize the 2CK FP,\cite{pang:91, affl:92, sela:11, saka:90,saka:92,izum:00} rendering it unstable on the lowest temperature ($T$) scales. The system instead flows ultimately to a stable strong coupling (SC), Fermi liquid-like FP below some characteristic low-temperature Fermi liquid scale.

For this reason, two-channel Kondo has experimentally been the most elusive of the various Kondo effects (we know of only one example\cite{poto:07} where it is believed to have been observed cleanly).
Potential observation of it relies on the fact that if inter-lead charge transfer
is sufficiently small, then a $T$-window can at least in principle exist over which the system flows close to the now-unstable 2CK FP -- such that non-Fermi liquid behavior could be observed -- before ultimately crossing over to the stable 
SC FP. The obvious questions then are:\cite{male:10} under what conditions does this arise, and are the resulting temperatures experimentally credible?

These questions are considered in the present work where, for vanishing magnetic field in the first instance (sec.\ref{zerofield}), we study directly the 2AIM, together with the effective low-energy models derived from  it (under Schrieffer-Wolff) to $\mathrm{2^{nd}}$ and 
$\mathrm{3^{rd}}$ order in the tunnel couplings; respectively the 2IKM, and a spin-model containing the key effects of \emph{cotunneling} inter-lead charge transfer. We also consider for comparison the model studied in
ref.\onlinecite{male:10}, in which solely \emph{direct} inter-lead charge transfer is added to the 2IKM.
The models themselves are specified in sec. \ref{section:models}, and
the numerical renormalization group (NRG) method\cite{wils:75, kris:80, kris2:80} (see ref. \onlinecite{bull:08} for a 
review) is employed to study them,
backed up by physical arguments.

In sec. \ref{magfield} we consider these models when a non-zero magnetic field is applied to the dots. 
For the channel-symmetric 2IKM in particular, we show that its zero-field quantum phase transition 
is the terminal endpoint of a line of transitions characterized by a 2CK FP. This raises the possibility
that, in the presence of sufficiently small inter-lead charge transfer, 2CK physics might be accessible by 
tuning a magnetic field; which question is then considered.
The paper ends with concluding remarks.


\section{Models}
\label{section:models}
We begin by specifying the models considered for TCDQDs, starting with the two-impurity Anderson model (2AIM)\cite{alex:64,gott:64,yama:79,jaya:82} as the canonical model for such.

\subsection{Two-impurity Anderson model}
\begin{figure}
\begin{center}
\includegraphics[height=35mm]{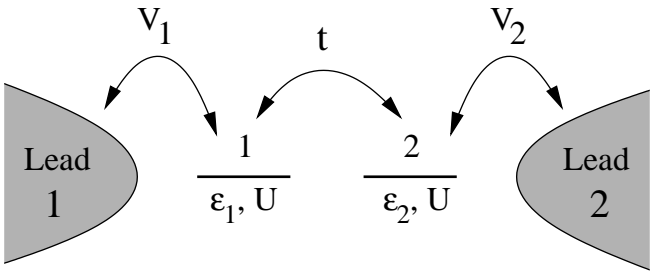}
\caption{Schematic of the 2AIM,
as discussed in text.
}
\label{fig:model}
\end{center}
\end{figure}
The 2AIM is illustrated schematically in Fig. \ref{fig:model}. It consists of two single-level dots
(labeled $\nu =1,2$, with level energies $\epsilon_{\nu}$ and on-level Coulomb replusion $U$), mutually tunnel-coupled by a hopping matrix element $t$; and with 
each dot tunnel-coupled to a separate non-interacting  metallic lead.
The model Hamiltonian is
\begin{equation}
\hat{H}_{\mathrm{2AIM}} = \hat{H}_{\mathrm{leads}} + \hat{H}_{\mathrm{dots}} + \hat{H}_{\mathrm{hyb}},
\end{equation}
where for the double quantum dot itself,
\begin{equation}
\begin{split}
\hat{H}_{\mathrm{dots}} =& \hat{H}_{\epsilon} + \hat{H}_{U} + \hat{H}_{t} \\
= &\sum_{\nu, \sigma}\epsilon_{\nu}\hat{n}_{\nu \sigma} + U\sum_{\nu}  \hat{n}_{\nu \uparrow}\hat{n}_{\nu \downarrow} + t\sum_{\sigma} \left({d}^{\dag}_{1\sigma}{d}^{\phantom{\dag}}_{2 \sigma} + \mathrm{H.c.} \right) 
\end{split}
\end{equation}
with $\hat{n}_{\nu\sigma}=d^{\dagger}_{\nu\sigma}d^{\phantom\dagger}_{\nu\sigma}$ the $\sigma =\uparrow/\downarrow$-spin number operator for dot $\nu$. For the two equivalent leads (likewise 
denoted $\nu =1,2$), 
\begin{equation}
\hat{H}_{\mathrm{leads}} = \sum_{\nu,\mathbf{k},\sigma} \epsilon_{\mathbf{k}}^{\phantom\dagger} c^{\dag}_{\nu \mathbf{k}\sigma}c^{\phantom{\dag}}_{\nu \mathbf{k}\sigma},
\end{equation}
and we consider the standard case\cite{hews:93} of a flat-band lead  with a (uniform) density of states per orbital of $\rho =1/(2D)$, with $D$ the half-bandwidth; denoting the total density of states by $\rho_{T} =N\rho$, with $N$ ($\rightarrow \infty$) the number of orbitals in a lead.
The hybridization term coupling the dots and leads is
\begin{equation}
\begin{split}
\hat{H}_{\mathrm{hyb}} = &\sum_{\nu, \mathbf{k},\sigma} V_{\nu}^{\phantom\dagger} \left({d}^{\dag}_{\nu \sigma} c^{\phantom{\dag}}_{\nu \mathbf{k}\sigma} + \mathrm{H.c.}\right)\\
=& \sum_{\nu,\sigma} \sqrt{N}V_{\nu}^{\phantom\dagger} \left(d^{\dagger}_{\nu\sigma}f_{\nu\sigma}^{\phantom\dagger} + \mathrm{H.c.}\right)
\end{split}
\end{equation}
such that dot-$\nu$ is tunnel-coupled to lead-$\nu$ with matrix element $V_{\nu}$; and where
\begin{equation}
f^{\dagger}_{\nu\sigma} = \frac{1}{\sqrt{N}} \sum_{\mathbf{k}} c_{\nu\mathbf{k}\sigma}^{\dagger}
\end{equation}
is the creation operator for the `$0$'-orbital of the Wilson chain\cite{wils:75,hews:93} for lead $\nu$.
Tunnel-coupling to lead-$\nu$ is embodied in the hybridization strength $\Gamma_{\nu} =\pi\rho_{T}V_{\nu}^{2}$. Unless explicitly stated otherwise, we consider the case of symmetric tunnel-coupling, $\Gamma_{1}=\Gamma_{2} =\Gamma$; and the zero-bias Fermi level of the leads, $E_{F}$, is taken as the zero of energy.

The effect of a magnetic field applied to the dots, which is considered in sec.~\ref{magfield}, is encompassed by including
\begin{equation}
\hat{H}_B =  -h \hat{S}_z
\label{BHam}
\end{equation}
with $\hat{S}_z =\tfrac{1}{2}\sum_{\nu}(\hat{n}_{\nu\uparrow}-\hat{n}_{\nu\downarrow})$
and $h=g \mu_B B$; where $B$ is the applied magnetic field and
$g$ the electron $g$-factor.

\subsubsection{Symmetries}
If $\epsilon_1 = \epsilon_2$ and $\Gamma_{1}=\Gamma_{2}=\Gamma$ (i.e.\ $V_1 = V_2$), the model is `left-right (LR) symmetric', meaning invariant under the 
transformation $d_{1\sigma}\leftrightarrow d_{2\sigma}$ and $c_{1\mathbf{k}\sigma} \leftrightarrow c_{2\mathbf{k}\sigma}$. In addition,  if $\epsilon_1 = \epsilon_2 = -U/2$ (but regardless of whether or not $\Gamma_{1} =\Gamma_{2}$), then the model is particle-hole (ph) symmetric, i.e. invariant under the ph transformation
$d^{\dagger}_{1\sigma} \leftrightarrow d^{\phantom{\dagger}}_{1\sigma}, d^{\dagger}_{2\sigma} \leftrightarrow - d^{\phantom{\dagger}}_{2\sigma}$, 
$c^{\dagger}_{1\mathbf{k}\sigma} \leftrightarrow c^{\phantom{\dagger}}_{1-\mathbf{k}\sigma}$ and $c^{\dagger}_{2\mathbf{k}\sigma} \leftrightarrow - c^{\phantom{\dagger}}_{2-\mathbf{k}\sigma}$.
In this paper we consider explicitly the 2AIM 
at ph symmetry.\cite{note:phsymm}
With LR symmetry also present, the full set of `bare' parameters for the 2AIM is simply $U/\Gamma$, $t/\Gamma$ and $D/\Gamma$. The bandwidth $D$ is naturally taken to be the largest energy scale in the problem, and for our NRG calculations in practice we take $D/\Gamma =100$.

\subsection{Schrieffer-Wolff transformations}
\label{SW}
The 2AIM, allowing as it does for charge fluctuations on the dots, exhibits a rich range of behavior across its full parameter space. Here we focus exclusively on the regime where each dot level is in essence singly occupied, as occurs for $U \gg t, \Gamma_1, \Gamma_2$ in the ph symmetric systems considered. In this case a 
Schrieffer-Wolff (SW) transformation\cite{schr:66} may be used to obtain an effective low-energy model for the system. 
This
involves perturbation theory in the tunnel couplings $V_1,V_2$ and $t$, i.e. the perturbing Hamiltonian is taken to be $\hat{H}_1 = \hat{H}_{\mathrm{hyb}} + \hat{H}_t$, and the only states of $\hat{H}_0 = \hat{H} - \hat{H}_1$ retained are those in which the dots are singly occupied (with a local unity operator denoted $\hat{1}^{\prime}$).

This perturbation theory can in principle be carried out to any order in $\hat{H}_{1}$. The leading non-vanishing contributions to the effective Hamiltonian arise to second order (specifically $\hat{1}' \hat{H}_1 (\epsilon_0 - \hat{H}_0)^{-1} \hat{P} \hat{H}_1\hat{1}'$, where $\hat{P} = \hat{1} - \hat{1}'$ is a projector and $\epsilon_0$ is the ground state energy of $H_0$); and
the effective low-energy model resulting from the second-order SW transformation on the 2AIM (neglecting retardation as usual\cite{hews:93}) is the much-studied two-impurity Kondo model (2IKM).\cite{jone:87,jone:88,jone:89,jone2:89,jone:91,affl:95} It consists of two spins-$\tfrac{1}{2}$, each coupled to a separate lead by antiferromagnetic (AF) Kondo 
couplings $J_{1}$ and $J_{2}$, and mutually coupled by an AF \emph{exchange} coupling $J$ -- precluding as such charge transfer between the leads.
The Hamiltonian is
\begin{equation}
\label{2ikm}
\hat{H}_{\mathrm{2IKM}} = J_{1} \hat{\textbf{S}}_1 \cdot \hat{\textbf{S}}_{01} + J_{2} \hat{\textbf{S}}_2 \cdot \hat{\textbf{S}}_{02} + J \hat{\textbf{S}}_1 \cdot \hat{\textbf{S}}_2 + \hat{H}_{\mathrm{leads}}
\end{equation}
where $\hat{\textbf{S}}_{\nu}$
is a spin-$\half$ operator representing dot $\nu =1,2$, and $\hat{\mathbf{S}}_{0\nu}$ is the spin-$\tfrac{1}{2}$ operator corresponding to the local spin density of lead $\nu$ ($\hat{\mathbf{S}}_{0\nu}=\sum_{\sigma ,\sigma^{\prime}}f^{\dagger}_{\nu\sigma}\boldsymbol{\sigma}^{\phantom\dagger}_{\sigma\sigma^{\prime}}f^{\phantom\dagger}_{\nu\sigma^{\prime}}$ with  $\boldsymbol{\sigma}$ the Pauli matrices).
From the SW transformation, the parameters of the 2IKM are related to those of the 2AIM by
$\rho J_{\nu} = 8\Gamma_{\nu}/\pi U$ and $J = 4t^2/U$.

It is however obvious that a \emph{second}-order SW transformation does not capture adequately the low-energy physics of the 2AIM, for it lacks the inter-lead 
charge-transfer processes that ensure the ground state of the 2AIM is always a Fermi liquid;
and which are central in understanding the role of 2CK physics in TCDQDs. 

As mentioned above, SW to higher orders can be carried out, and to capture inter-lead cotunneling charge transfer one must go to third order.
The additional third-order term arising from
a SW transformation of the 2AIM,
$\hat{H}_3 = \hat{1}' \hat{H}_1 (\epsilon_0 - \hat{H}_0)^{-1} \hat{P} \hat{H}_1 (\epsilon_0 - \hat{H}_0)^{-1} \hat{P} \hat{H}_1\hat{1}'$, is given after 
lengthy calculation by
\begin{equation}
\label{H3}
\begin{split}
\hat{H}_3 = V_{LR}\left[\sum_{\sigma}(f^{\dag}_{1\sigma}f^{\phantom{\dag}}_{2\sigma}+f^{\dag}_{2\sigma}   f^{\phantom{\dag}}_{1\sigma})\hat{\textbf{S}}_1 \cdot \hat{\textbf{S}}_2 + 2\hat{\textbf{A}} \cdot (\hat{\textbf{S}}_1 \times \hat{\textbf{S}}_2)  \right]
\end{split}
\end{equation}
where 
$V_{LR} =(16t V_{1}V_{2})/U^{2}$ $=\sqrt{(J J_{1}J_{2})/U}$ and
$\hat{\textbf{A}}$ is a vector operator with components
\begin{equation}
\hat{\textbf{A}}= i \sum_{\sigma,\sigma^{\prime}} \left( f_{1\sigma}^{\dagger}\boldsymbol{\sigma}^{\phantom\dagger}_{\sigma\sigma^{\prime}}f^{\phantom\dagger}_{2\sigma^{\prime}}-f_{2\sigma}^{\dagger}\boldsymbol{\sigma}^{\phantom\dagger}_{\sigma\sigma^{\prime}}f^{\phantom\dagger}_{1\sigma^{\prime}}\right)
\end{equation}
Note that $\hat{\textbf{A}}$, which is self-adjoint and odd under $1\leftrightarrow 2$ exchange, is not a spin operator; its components satisfying the commutation relations $[\hat{A}^{\alpha},\hat{A}^{\beta}] = i \epsilon_{\alpha\beta\gamma}\hat{S}_{0}^{\gamma}$, where $\hat{\textbf{S}}_{0}=\hat{\textbf{S}}_{01}+\hat{\textbf{S}}_{02}$ (with $\alpha,\beta,\gamma \in (x,y,z)$ and $\epsilon_{\alpha\beta\gamma}$ the Levi-Civita symbol).
$\hat{H}_3$, in which charge transfer between the leads is mediated by the dot spins,  
is clearly a rather complicated object.
It was 
obtained previously in Ref. \onlinecite{sela3:09}, but subsequently neglected. In the following, we refer to
the third-order effective low-energy model specified by $ \hat{H}_{\mathrm{2IKM}}+\hat{H}_3$ as the $H_{3}$ model.

It is important to emphasize 
that the charge-transfer processes in the 2AIM involve cotunneling, i.e.
are mediated by the dot spin degrees-of-freedom. In recent work\cite{sela:09, male:10}, the 2IKM with an additional \emph{direct} lead-lead tunneling term was studied as a model for a TCDQD. The Hamiltonian 
considered was $\hat{H}_{\mathrm{2IKM}}+\hat{H}'_{3}$, with 
a charge transfer term
\begin{equation}
\hat{H}'_{3} = V'_{LR}\sum_{\sigma}(f^{\dag}_{1\sigma}f^{\phantom{\dag}}_{2\sigma}+f^{\dag}_{2\sigma}   f^{\phantom{\dag}}_{1\sigma})
\label{H'3}
\end{equation}
where $V'_{LR} = \tfrac{1}{4}V_{LR}$. We refer to this as the $H_{3}^{\prime}$ model.
It is \emph{not} the correct effective low-energy model for the 2AIM, and as such should not be expected to exhibit the same physics as the 2AIM even at low energies
(indeed it does not\cite{male:10}). 
We include it here purely for comparison to the 2AIM and $H_{3}$ models, to illustrate that adding the $\hat{H}_3$ term to the 2IKM has a notably different effect to adding the $\hat{H}'_3$ term.

We have now looked at all the models to be considered in this paper: the full 2AIM, the effective low-energy models for the 2AIM derived by SW transformation to 2nd and 3rd order (the 2IKM and $H_3$ models respectively), and the $H'_3$ model.
In the following section we use the NRG\cite{wils:75, kris:80, kris2:80, bull:08,pete:06, weic:07} to obtain results for these models  at zero field. We typically retain between 2000 and 4000 states at each NRG iteration, and use an NRG discretization parameter $\Lambda = 3$.

\section{2CK physics at zero magnetic field}
\label{zerofield}
We begin with a brief summary of the 
2IKM at zero field, before considering the effect of inter-lead charge transfer as included in the 2AIM, $H_3$ and $H'_3$ models.

\subsection{Two-impurity Kondo model}   

It is well known\cite{jone:88, jone:89, jone2:89, affl:92, affl:95} that the LR-symmetric 2IKM ($J_{1}=J_{2}$)
exhibits a quantum phase transition (QPT) at a critical value $J_{c}$ of the inter-spin exchange 
$J$; with $J_{c} \sim {\cal{O}}(\tk)$, and $\tk$ the Kondo scale of the system when the spins are decoupled, $J=0$.
The transition separates a local singlet (LS) phase arising for $J>J_{c}$, in which the two spins bind to form a singlet, from 
a phase in which each spin is separately quenched by Kondo coupling to its attached lead
(we refer to it as the `Kondo singlet' (KS) phase).
These phases are readily identified from the phase shift, $\delta_e$, in the even combination of conduction channels/leads,
 which vanishes in the LS phase, and 
is $\pi/2$ in the KS phase (see Refs.\onlinecite{affl:95, geor:99, male:10} for details).

The 
FP for the transition is distinct from those of the LS or KS phases, and corresponds 
to the 2CK FP,\cite{jone:88, jone:89,affl:92} as known e.g.\ from conformal field theory\cite{affl:92,affl:95,gan:95} and NRG\cite{affl:95} studies (albeit that the operator content and finite size spectrum of the critical FP differ slightly from the 2CK FP\cite{affl:95}). 
On decreasing the temperature ($T$)/energy scale at $J = J_c$, the system flows 
from a Local Moment (LM) FP -- where the dot spins are effectively decoupled from each other and from the leads,
with a corresponding entropy $S_{\mathrm{imp}} =\ln 4$ 
-- to the critical 2CK FP characterized by $S_{\mathrm{imp}} =\tfrac{1}{2}\ln 2$, 
on the scale $T\sim \tk$ (so that $\tk$ is also in effect the two-channel Kondo scale).

We comment in passing on the relation\cite{jone:88,jone:89,jone2:89,jone:91, geor:99} $J_{c} =\alpha \tk$ with $\alpha$ a constant,
which we find from NRG calculations indeed holds for sufficiently small $\rho J_{1} \ll 1$.\cite{note:alpha}
The precise value of $\alpha$ naturally depends on how $\tk$ (pertaining to $J=0$) is defined; and in this there is 
freedom of choice. In practice we choose $\tk$ to be the $T$ for which $S_{\mathrm{imp}}(\tk) =\ln 2$, halfway between $S_{\mathrm{imp}} =\ln 4$ characteristic of the LM FP and the $T=0$ entropy $S_{\mathrm{imp}} =0$ 
for the stable strong coupling FP in the KS phase. With this, the constant $\alpha \approx 8$. If instead we had chosen to define $\tk$ as $8/2.2$ times the temperature for which $S_{\mathrm{imp}}=\ln 2$, then $\alpha \simeq 2.2$; as often quoted in the
literature.\cite{jone:88,jone:89,jone2:89,jone:91, geor:99} There is however no fundamental distinction between these
different practical definitions of $\tk$.

Finally, while the comments above refer to the LR-symmetric case,
we add that the QPT is also  known\cite{zara:06,mitc:09} to remain
robust to $J_{1}\neq J_{2}$, with a line of 2CK critical FPs in the ($J_{1},J_{2}$)-plane separating LS and KS phases; and a critical $J_{c}$ dependent on $\tk^{(1)}$ and $\tk^{(2)}$, the two distinct $J=0$ Kondo scales now arising.

\subsection{Effects of charge transfer}
\label{CT}
We now look at the effect of adding inter-lead charge transfer processes to the 2IKM (focussing on the LR-symmetric case).
These destroy the QPT occurring in the 2IKM, and with it the stability of the 2CK quantum critical point; the pristine transition being replaced by a continuous crossover between KS and LS ground states, characterized by a stable SC FP with $S_{\mathrm{imp}}(T=0)=0$. 
This is well known for the 2AIM\cite{saka:90,saka:92,izum:00} and $H_{3}^{\prime}$\cite{sela:11,male:10} models; and 
our NRG calculations indicate the same for the $H_{3}$ model
(unsurprisingly, it being the effective low-energy model for the 2AIM).

Although the 2CK FP is rendered unstable by charge transfer, with decreasing $T$ the system may first flow close to it on a scale $T \sim \tk$ (as for the 2IKM) 
-- evident e.g. in a characteristic $\tfrac{1}{2}\ln 2$ entropy plateau  -- before flowing to the stable SC FP on a low-energy Fermi liquid scale $T^{*}$ [which we calculate in practice from $S_{\mathrm{imp}}(T^{*})=\tfrac{1}{4}\ln 2$, halfway between the characteristic values for the 2CK FP and stable SC FP]. \emph{If} this situation arises, then the 2CK FP is effectively `visible' at finite temperature, occurring over an appreciable $T$-window provided
\begin{equation}
T^{*} \ll T \ll \tk.
\label{condition}
\end{equation}
The obvious questions then are\cite{male:10}: under what conditions does this behavior arise? And for the 2AIM in particular
(as the canonical model for TCDQDs), does it occur for experimentally realistic temperatures?
\begin{figure}[t]
\begin{center}
\includegraphics[height=45mm]{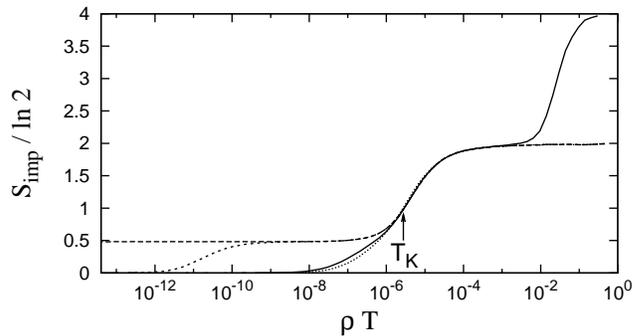}
\caption{$S_{\mathrm{imp}}$ \emph{vs} $\rho T$ ($\rho =1/(2D)$) from NRG for: 2AIM with $U/\Gamma = 20$ and $t=t_c$ (solid line), and for the 2IKM (long-dashed), $H_3$ (dotted) and $H'_3$ (short-dashed line) models with $\rho J_1 = 0.093$ and $J = J_c$. 
The common Kondo scale $\tk$ is indicated by an arrow.
}
\label{fig:B0_entropyComparison}
\end{center}
\end{figure}

To answer these questions we consider explicitly the $T$-dependence of the entropy $S_{\mathrm{imp}}(T)$. As for the 2IKM,  $\tk$ is the Kondo scale when the spins/dots are decoupled, viz. $J$=0 for the spin-models $H_{3}$ and $H_{3}^{\prime}$, and $t=0$ for the 2AIM; with $\tk$ defined 
via $S_{\mathrm{imp}}(\tk) = \ln 2$, as above (although in practice the resultant $\tk$ differs insignificantly from that which can be read off e.g. fig. \ref{fig:B0_entropyComparison} below from $S_{\mathrm{imp}}(\tk) = \ln 2$). To optimise the possibility of observing the 2CK FP at finite-$T$ we follow ref.~\onlinecite{male:10} and consider $J=J_{c}$ ($\sim \tk$) for all spin models and $t=t_{c}$ for the 2AIM -- where the models flow closest to the 2CK FP -- chosen\cite{male:10} in either case so that the even channel phase shift $\delta_{e} =\pi/4$. The phase shifts $\delta_{e}$ are themselves  determined straightforwardly from the potential scattering on the even lead at the SC FP\cite{kris2:80} (itself obtained by comparing the NRG FP energy levels with those calculated separately from free even and odd conduction chains with equal and opposite potential scattering).

Fig. \ref{fig:B0_entropyComparison} shows NRG results for $S_{\mathrm{imp}}(T)$ for the 2AIM with $U/\Gamma =20$ (solid line), with $\tk$ as indicated. We wish to compare this to spin-models ($H_{3}$, $H_{3}^{\prime}$ and 2IKM) with the same $\tk$, to which end we consider
$\rho J_{1}=\rho J_{2} = 0.093$ (which value differs somewhat from that given by the SW transformation, reflecting simply the fact that for $U/\Gamma =20$ the SW asymptotics for $\rho J_1$ have not quite been reached). And for the spin models with charge transfer, 
we take $V_{LR}=\sqrt{J_{c}J_{1}^{2}/U}$ as in sec.~\ref{SW}.

On decreasing $T$ for the 2IKM with $J=J_{c}$, the system naturally flows to the 2CK FP ($S_{\mathrm{imp}}=\tfrac{1}{2}\ln 2$) for $T \sim \tk$, and remains there. For the $H_{3}^{\prime}$ model, with direct inter-lead tunneling, there is a clear $\tfrac{1}{2}\ln 2$ entropy plateau before flow to the stable SC FP ($S_{\mathrm{imp}}=0$) on the scale $T \sim T^{*}$; with $T^{*} \ll \tk$ in this case, so that 2CK physics arises over an appreciable temperature window. The behavior of the 2AIM for $U/\Gamma =20$ is quite different. After descent on the scale $T \sim U$ from its trivial high-$T$ limit $S_{\mathrm{imp}} = \ln16$ to the LM FP with $S_{\mathrm{imp}} = \ln 4$, the system flows \emph{directly} to the SC FP on the scale $T \sim \tk$; with no hint of flow in the vicinity of the 2CK FP. The $H_{3}$ model is also seen to exhibit the same low-energy behavior.

\begin{figure}[t]
\begin{center}
\includegraphics[height=45mm]{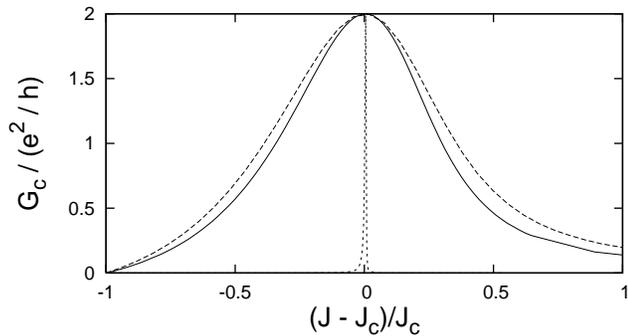}
\caption{For the same parameters as Fig. \ref{fig:B0_entropyComparison}, the 
$T=0$ zero-bias conductance $G_{c}$ \emph{vs} $(J-J_{c})/J_{c}$ for the 2AIM (solid line), the $H_3$ model (long-dashed) and the $H_3'$ model (short-dashed). 
}
\label{fig:cond}
\end{center}
\end{figure}

NRG may also be used to calculate the $T=0$ zero-bias conductance, $G_c$. For LR-symmetric systems,\cite{geor:99}
\begin{equation}
G_c = \frac{2e^2}{h} \sin^2 (\delta_e - \delta_o) =  \frac{2e^2}{h} \sin^2 (2\delta_e)
\end{equation}
with $\delta_{e(o)}$ the phase shift in the even (odd) channel,
and the second equality follows at ph symmetry.\cite{affl:95} Calculation of $\delta_{e}$ thus gives $G_{c}$ directly. Fig. \ref{fig:cond} shows $G_c$ \emph{vs} $[J-J_{c}]/J_{c}$ for the 2AIM (where $J=4t^{2}/U$ is taken), and the $H_{3}$ and $H_{3}^{\prime}$ models; with $J_{c} \sim \tk$ as ever. For the $H_{3}^{\prime}$ model the halfwidth of this conductance peak (\emph{vs} $[J-J_{c}]/J_{c}$) is known\cite{male:10} to be proportional to $\sqrt{T^{*}/\tk}$; the fact that it is evidently $\ll 1$ indicating the clear scale separation  $T^{*} \ll \tk$ seen from the entropies of fig.\ref{fig:B0_entropyComparison}. $G_{c}$ for the 2AIM and $H_{3}$ models are similar, as expected (the differences again reflect that SW is asymptotically exact only as $U/\Gamma \rightarrow \infty$). In these cases, by contrast,
the conductance halfwidth is clearly ${\cal{O}}(1)$. This too is consonant with the entropies shown in fig. \ref{fig:B0_entropyComparison}, where for $U/\Gamma =20$ there is no flow in the vicinity of the 2CK FP, and as such $\tk$ is the sole low-energy scale in the problem.

We also point out here that our conductance results are in agreement both with previous work on the 2AIM\cite{izum:00}, and recent work on the $H_3'$ model\cite{male:10}: that the two models give different results is a natural consequence of the fact that they are not simply related by SW transformation. \\

\begin{figure}
\begin{center}
\includegraphics[height=45mm]{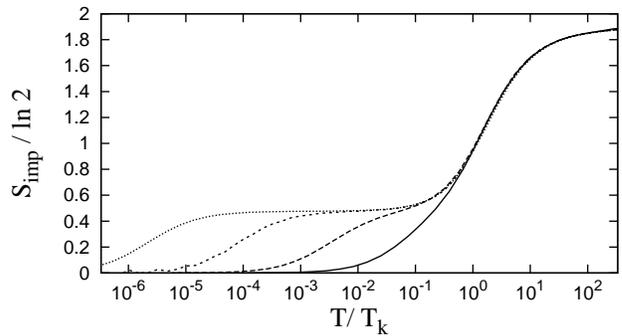}
\caption{$S_{\mathrm{imp}}(T)$ \emph{vs} $T/\tk$ for the 2AIM (with $t = t_c$) for $U/\Gamma = 20$ (solid line), $30$ (long dash), $40$ (short dash), $50$ (dotted). A $\half \ln 2$ plateau appears as $U/\Gamma$ is increased, indicating flow in the vicinity of the 2CK FP.
}
\label{fig:B0_varU_entropy}
\end{center}
\end{figure}

 The 2CK FP, as manifest in a $\tfrac{1}{2}\ln2$ entropy plateau, does not then occur for $U/\Gamma =20$ in the 2AIM (or in its effective low-energy model, $H_{3}$). To observe it in the 2AIM necessitates a larger $U/\Gamma$ in order to suppress inter-lead cotunneling charge transfer (i.e. to reduce $V_{LR}$ in the effective low-energy  model), although this will also reduce $\tk$ itself (since\cite{hews:93} $\tk \propto \exp(-\pi U/(8\Gamma))$). To illustrate this, 
Fig. \ref{fig:B0_varU_entropy} shows $S_{\mathrm{imp}}(T)$ \emph{vs} $T/\tk$ for the 2AIM,
for various values of $U/\Gamma$. On increasing the interaction
a $\half \ln 2$ entropy plateau appears, indicating the opening of a temperature window in which the system flows in the vicinity  of the 2CK FP, with a clear scale separation $T^{*} \ll \tk$ for sufficiently large $U/\Gamma$, in practice $U/\Gamma \gtrsim 40$. 
In this regime our numerics  are consistent with the 
form
\begin{equation}
\frac{T^{*}}{\tk} = F\left(\frac{U}{\Gamma}\right)~\frac{\tk}{U},
\end{equation}
and although we have not performed exhaustive calculations our results indicate $F(x) \sim b x^{2}$ with
$b \sim {\cal{O}}(1)$ a constant.
The behavior $T^* \propto \tk^{2}$ also arises\cite{male:10} in the $H_{3}^{\prime}$ model.\cite{note:Tstar}
Here it is found\cite{male:10} that 
\begin{equation}
T^* = b' (\rho V'_{LR})^2 \tk
\label{affleckT*}
\end{equation}
with $b^{\prime} \sim 10^{2}$ approximately constant, $V^{\prime}_{LR} =\tfrac{1}{4}V_{LR}$ and (as above)
$(\rho V_{LR})^{2} = J_{c} (\rho J_{1})^{2}/U$.  Since $J_{c}\sim \tk$ itself, eq.\ref{affleckT*} gives
$T^* \propto \tk^{2}$.

As mentioned above, increasing $U/\Gamma$ in the 2AIM in order to access a reasonable $T$-window for 2CK physics, naturally has the effect of reducing $\tk$, the minimum temperature needed to access the 2CK regime. From fig. \ref{fig:B0_varU_entropy}, $U/\Gamma \gtrsim 40$ is in practice required for a reasonable window to arise. But for
$U/\Gamma =40$, our NRG results give $\tk/U \sim 4 \times 10^{-9}$. Taking $U = 2$ meV, as is experimentally typical\cite{inge:08,jeon:01, grab:06, chan:09}, this corresponds to $\tk \sim  10^{-7}\mathrm{K}$ -- far lower than can be reached in experiment (a minimum of around $10 \mathrm{mK}$). Even reducing $U/\Gamma$ to $30$, where a $\tfrac{1}{2}\ln 2$ `plateau' is just about visible in fig. \ref{fig:B0_varU_entropy}, yields 
$\tk \sim  10^{-5}\mathrm{K}$, some three orders of magnitude lower than what is 
experimentally feasible. In contrast to previous estimates,\cite{male:10} our conclusion is that it is highly unlikely 2CK physics could realistically be observed in tunnel-coupled DQDs.\\

 While we have focussed above on LR-symmetric systems, the physics of the models considered remains qualitatively the same when LR symmetry is broken:
inter-lead charge transfer, whether of cotunneling or `direct' form, destroys the
QPT exhibited by the 2IKM. If we consider breaking LR symmetry by decreasing just one of the dot-lead couplings, then inter-lead charge transfer is suppressed, leading to a decrease in the $T^*$ scale. This 
raises the question whether 2CK physics might more readily be observed in a strongly 
asymmetric TCDQD device. That is not however the case.
As for the LR-asymmetric 2IKM, the (experimentally relevant) 
2AIM, if it flows at all in the vicinity of the 2CK FP, does so on the scale
of $\mathrm{min}(\tk^{(1)}, \tk^{(2)})$. Decreasing e.g. $\Gamma_2$ at fixed $\Gamma_1$ will thus not only reduce the $T^*$ scale, 
but also the scale $\tk^{(2)}$ on which the system flows to the 2CK FP. Introducing LR-asymmetry in this way therefore decreases the temperature at which 2CK physics might be observed.


\section{2CK physics at finite magnetic field}
\label{magfield}
Having discussed the possibility of observing 2CK physics in a TCDQD at zero magnetic field, we now ask
whether it might be possible to do so 
at finite field. Again we begin with the 2IKM, showing first that in the LR-symmetric case the quantum phase transition known to arise at zero-field is the terminal point of a line of QPTs accessed by tuning the magnetic field.

\subsection{Two-impurity Kondo model}
At zero field, the trivial atomic-limit of the 2IKM (i.e. eq.\eqref{2ikm} with $J_1 = J_2 = 0$) has a singlet ground state, $\ket{S} = \frac{1}{\sqrt{2}}(\ket{\uparrow \downarrow} - \ket{\downarrow \uparrow})$ (representing the two dot spins), with a degenerate triplet at energy $J$ above the ground state.  Application of a magnetic field to the dot spins (eq. \ref{BHam}) lowers the energy of one triplet component, 
$\ket{T_1} = \ket{\uparrow \uparrow}$ (for $h > 0$), 
with a triplet-singlet energy difference  $E_{T_{1}}-E_{S} = (J-h)$. 
At a `critical' field 
$h = h_c = J$ the ground state will thus be doubly degenerate, constituting as such as pseudospin-$\half$ comprised of $\ket{S}$ and $\ket{T_1}$.

The energies of the remaining $\ket{T_{0}}$ and $\ket{T_{-1}}$ triplet components lie at least $J$ above the ground state. On coupling the dot spins to the leads, only the pseudospin-$\half$ need therefore be retained in the low-energy manifold of dot states, provided $J \gg \tk \sim J_{c}$ (or $J \gg \mathrm{max}(\tk^{(1)},\tk^{(2)})$ for $J_{1} \neq J_{2}$); since $\tk$ (defined as usual for $J=0$) is the sole low-energy scale in the problem when $J=0$.
In otherwords, the $\ket{T_{0}}$ and $\ket{T_{-1}}$ triplet components may be neglected provided $J \gg J_{c}$, and only the pseudospin-$\half$ need be retained. Since this pseudospin can be flipped by the Kondo exchange on coupling to the leads, and since it is coupled to two leads, we thus expect\cite{kiko:07} that two-channel Kondo physics should arise in the 2IKM with $J \gg J_{c}$, at a critical field
$h=h_{c} \sim J$.

\begin{figure}[t]
\begin{center}
\includegraphics[height=45mm]{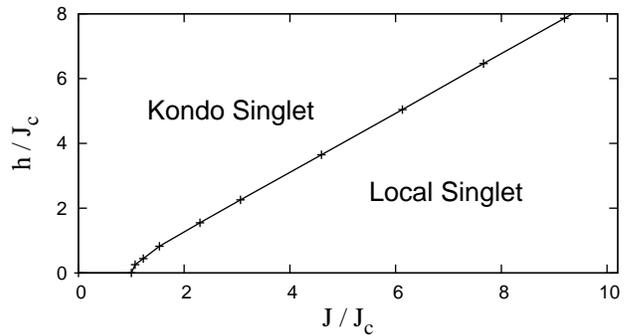}
\caption{NRG-determined phase diagram for LR-symmetric 2IKM as a function of $h$ and $J$, for $\rho J_1 = \rho J_2 = 0.093$. $J_{c}$ ($\sim \tk$) is the critical $J$ for zero field ($\rho J_{c} \simeq 1.63 \times 10^{-5}$ here). 
}
\label{fig:BJphase}
\end{center}
\end{figure}

The preceding physical argument may be put on firmer footing by deriving an effective low-energy model for the 2IKM with $J \gg J_c$ and $h \simeq h_{c}$,  retaining only the  states which form the two components of the pseudospin (with local unity operator $\hat{1}' = \ket{S}\bra{S} + \ket{T_1}\bra{T_1}$); i.e. from
$\hat{H}_{\mathrm{eff}}  =\hat{1}^{\prime}\hat{H}_{\mathrm{2IKM}}\hat{1}^{\prime}$ to leading order, with $\hat{H}_{\mathrm{2IKM}}$ in eq.\ref{2ikm}. This yields\cite{note:cantran} the following effective low-energy model
$\hat{H}_{\mathrm{eff}}=\hat{H}_{\mathrm{eff}}^{\prime}+\hat{O}$:
\begin{equation}
\begin{split}
\hat{H}_{\mathrm{eff}} &= \tfrac{1}{2}J_1\left[\hat{S}_{01}^z \hat{\tau}^z + \tfrac{1}{\sqrt{2}}(\hat{S}_{01}^+ \hat{\tau}^- + \hat{S}_{01}^- \hat{\tau}^+)\right]\\
&+ \tfrac{1}{2}J_2\left[\hat{S}_{02}^z \hat{\tau}^z + \tfrac{1}{\sqrt{2}}(\hat{S}_{02}^+ \hat{\tau}^- + \hat{S}_{02}^- \hat{\tau}^+)\right]\\
&+ \hat{O}
\end{split}
\label{heff}
\end{equation}
(in addition to $\hat{H}_{\mathrm{leads}}$, taken as read).
Here 
$\hat{\boldsymbol{\tau}}$ denotes the pseudospin-$\half$, with components $\hat{\tau}^z = \half(\ket{T_1}\bra{T_1} - \ket{S}\bra{S})$, $\hat{\tau}^+ = \ket{T_1}\bra{S}$ and $\hat{\tau}^{-} =(\hat{\tau}^{+})^{\dagger}$; and  $\hat{O}=\hat{O}_{1}+\hat{O}_{2}$ with
\begin{equation}
\begin{split}
\hat{O}_{1} =& \tfrac{1}{4} J_1\hat{S}_{01}^z + \tfrac{1}{4} J_2\hat{S}_{02}^z \\
\hat{O}_{2} =& (J-h)\hat{\tau}^{z}.
\end{split}
\label{O}
\end{equation}
The first two lines of $\hat{H}_{\mathrm{eff}}$, denoted $\hat{H}_{\mathrm{eff}}^{\prime}$, comprise a two-channel Kondo model with both channel anisotropy (for $J_{1} \neq J_{2}$), which is well known to be a relevant perturbation to the 2CK model;\cite{pang:91, affl:92, sela:11} and with spin anisotropy, known to be irrelevant to the 2CK FP.\cite{pang:91, affl2:92} 

Since any channel anisotropy destroys the 2CK FP, consider first the LR-symmetric 
case $J_{1}=J_{2}$. Now we must consider the effect of $\hat{O}=\hat{O}_{1}+\hat{O}_{2}$, each term of which is a relevant perturbation,\cite{affl2:92} separately rendering the 2CK FP unstable (as we have also confirmed directly via NRG on eq.\ref{heff}). However for any given $J$, $J_{1}$ and $J_{2}$, the magnetic field $h$ is a free parameter; which can then be tuned to ensure a vanishing \emph{coefficient} of the relevant primary field associated with $\hat{O}$, rendering it ineffective and the 2CK FP in consequence stable. This is the critical field, $h=h_{c}$.
From the physical arguments above we expect $h_{c} \sim J$ (although not identically $J$, as is obvious from eq.\ref{O}).

The above arguments imply that for the LR-symmetric 2IKM ($J_{1}=J_{2}$) with $J \gg J_{c}$, 
there should be a QPT at a critical $h_{c}$, with a 2CK critical FP; and hence a line of 2CK critical FPs in the ($h, J$)-plane.
We have confirmed this 
with NRG calculations on the 2IKM. Fig. \ref{fig:BJphase}
shows an illustrative phase diagram for fixed $\rho J_{1}=\rho J_{2}$ as a function of ($h/J_{c}, J/J_{c}$) (recall that $J_{c}$ is the critical $J$ for zero-field, with $J_{c} \sim \tk$); the critical line $h_{c}(J/J_{c})$ separating an $h>h_{c}$ phase which is continuously connected to the zero-field KS state, from that connected to the zero-field LS phase.
Although the arguments given above apply strictly to $J \gg J_{c}$, the transition is seen to extend continuously down to $J=J_{c}$ (as a simple appeal to continuity would suggest). As expected from the physical arguments above, $h_{c} \simeq J$ indeed arises for sufficiently large $J$. Indeed near-linear behavior is seen in practice to set in for $J/J_c\gtrsim 2$ or so, and for $J/J_c\gg1$ we have confirmed the linear form $h_c=-a+bJ$ where the gradient $b\to 1$ (and $a>0$ is a constant).
\begin{figure}[t]
\begin{center}
\includegraphics[height=45mm]{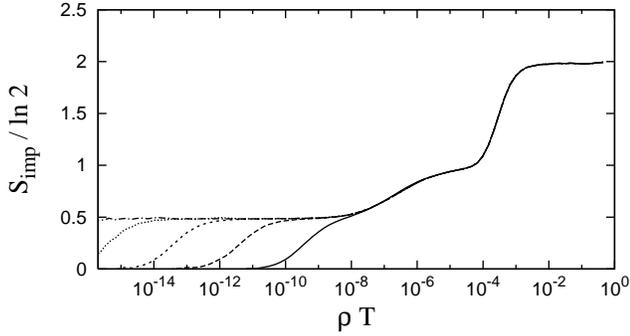}
\caption{$S_{\mathrm{imp}}(T)$ \emph{vs} $\rho T$ for 2IKM with $\rho J_{1}=0.093$ and $\rho J = 5 \times 10^{-4}$ 
($\sim 31 \rho J_{c} \gg \rho J_{c}$)
at the critical field $h=h_{c}$ ($\rho h_{c} = 4.55 \times 10^{-4}$ or $h_{c}/J = 0.91$), 
upon decreasing $J_{2}$ from $J_{1}$: $2\rho (J_{1}-J_{2}) =0$ (point-dash line), $10^{-6}$ (dotted), $10^{-5}$ short dash), $10^{-4}$ (long dash) and $10^{-3}$ (solid); corresponding respectively to $[J_{1}-J_{2}]/J_{c} = 0,
0.03, 0.3, 3$ and $30$.
}
\label{fig:Bfield_varJ2_entropy}
\end{center}
\end{figure}

For $J_1 \neq J_2$ by contrast, $\hat{H}_{\mathrm{eff}}$ (eq.\ref{heff}) is channel anisotropic.  The 2CK FP is consequently unstable,\cite{pang:91, affl:92, sela:11} as likewise follows from the arguments above
for the 2IKM at $h=h_{c}$ (at least for $J \gg J_{c}$). The system instead flows to a stable SC FP with $S_{\mathrm{imp}}(T=0)=0$; flowing for sufficiently small $|J_{1}-J_{2}|$ \emph{from} the 2CK 
critical 
FP ($S_{\mathrm{imp}}=\half\ln 2$) to the stable SC FP, on a scale $T_{*}$ known from the 2CK model\cite{2ck:BA_andrei_jerez,pang:91} to vanish as $T_{*} \sim (J_1 - J_2)^2$.
The validity of this picture has been confirmed by NRG and is illustrated in fig. \ref{fig:Bfield_varJ2_entropy}, showing the $T$-dependence of $S_{\mathrm{imp}}(T)$ for the 2IKM at the critical $h = h_c$ for fixed $J$ and $J_1$, upon decreasing  $J_2$ from 
$J_{2} =J_{1}$ where the 2CK critical FP is stable. 
The low-energy scale $T_{*}$ -- which in practice may be identified from $S_{\mathrm{imp}}(T_{*}) =\tfrac{1}{4}\ln2$ -- is immediately evident on increasing $J_{1}-J_{2}$ from zero; and analysis of the numerics indeed confirms it to 
vanish as $T_{*} \sim (J_1 - J_2)^2$. 

We also note that the temperature scale (call it $\tk^{\prime}$) on which the $\half\ln 2$ entropy plateau in 
fig. \ref{fig:Bfield_varJ2_entropy} is reached, is visibly lower than its counterpart shown in fig. \ref{fig:B0_entropyComparison} for the 2IKM at zero field (with the same $\rho J_{1}$), which is $\tk \propto \exp(-1/\rho J_{1})$.\cite{hews:93,ande:70} 
This can be understood from the low-energy model 
$\hat{H}^{\prime}_{\mathrm{eff}}$ (as appropriate to fig. \ref{fig:Bfield_varJ2_entropy}), it being sufficient
to consider the channel-symmetric case $J_{1}=J_{2}$.  This is a spin-anisotropic 2CK model, of form
$\hat{H}^{\prime}_{\mathrm{eff}} =\sum_{\nu =1,2}[ J_{z}\hat{S}^{z}_{0\nu}\hat{\tau}^{z} +\half J_{\perp}(\hat{S}^{+}_{0\nu}\hat{\tau}^{-}+\hat{S}^{-}_{0\nu}\hat{\tau}^{+})]$
with (see eq.\ref{heff}) exchange couplings $J_{z} =J_{1}/2$ and 
$J_{\perp} =J_{1}/\sqrt{2}$ (each less than $J_{1}$). And the characteristic low-energy scale for the model, $\tk^{\prime}$, on which temperature scale the $\half\ln 2$ entropy is approached, is readily shown from
perturbative scaling\cite{hews:93,ande:70} to be $\tk^{\prime} \propto \exp(-\tfrac{\pi}{2}\tfrac{1}{\rho J_{1}})$, whence $\tk^{\prime} \ll \tk$. 

This means, in otherwords, that much lower temperatures are needed to observe 2CK physics at the critical point of the 2IKM at finite field than at zero field -- scarcely a viable prospect in the light of sec.\ref{CT}.

\subsection{Two-impurity Anderson model}

\begin{figure}[t]
\begin{center}
\includegraphics[height=45mm]{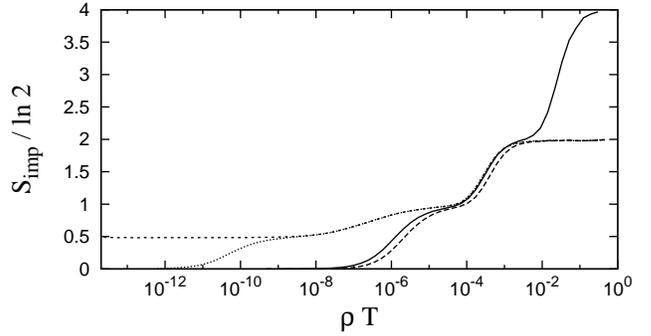}
\caption{$S_{\mathrm{imp}}(T)$ \emph{vs} $\rho T$ for different models with $h \simeq h_c$. Shown for
the 2IKM (short dashed line), $H'_3$ model (dotted line), and $H_3$ model (long dash); in all cases,
$\rho J_{1}=\rho J_{2}= 0.093$, $\rho J = 5\times 10^{-4}$ ($\gg \rho J_{c}$), and $V_{LR} = \sqrt{J J_1 J_2/U}$.
For the 2AIM (solid line), we consider $U/\Gamma =20$ and $t/\Gamma = 0.71$ (corresponding to $\rho J=\rho [4t^{2}/U] = 5\times 10^{-4}$).
}
\label{fig:Bfield_entropycomparison}
\end{center}
\end{figure}

We now turn to the 2AIM at finite field. If $t \gg t_c$ (i.e. $J = 4t^2/U \gg J_c$), we expect similar behavior to the 2IKM: a sufficiently large magnetic field $\sim h_{c}$ renders the singlet and lowest triplet atomic-limit states degenerate, suggesting the possibility of 2CK physics at low temperature. As was the case in sec.~\ref{CT}
however, whether two-channel Kondo can actually be observed in the 2AIM at finite field depends on the effect of  charge transfer inherent to the model, since this destabilises the 2CK FP.

Following the approach of sec.~\ref{CT}, we 
compare the behavior of the (channel-isotropic) 2AIM with two effective low-energy models: the $H_{3}'$ model in a magnetic field $h\sim h_c$, and the result of a third-order Schrieffer-Wolff transformation of the 2AIM. The latter is found to be equivalent to the $H_{3}$ model at $h\sim h_c$ upon retaining only the $|S\rangle$ and $|T_1\rangle$ dot states (\emph{viz} 
$\hat{H}^{\prime\prime}_{\mathrm{eff}} = \hat{1}^{\prime}(\hat{H}_{\mathrm{2IKM}}+\hat{H}_{3})\hat{1}^{\prime}$),
and is given explicitly by
\begin{equation}
\begin{split}
\hat{H}_{\mathrm{eff}}'' = \hat{H}_{\mathrm{eff}} - V_{LR}\Bigl[\sum_\sigma \sigma(f^{\dag}_{1\sigma}f^{\phantom\dag}_{2\sigma} + f^{\dag}_{2\sigma}f^{\phantom\dag}_{1\sigma})(\hat{\tau}_z - \tfrac{1}{4}) \\
+ \tfrac{1}{\sqrt{2}}(f^{\dag}_{1\uparrow}f^{\phantom\dag}_{2\downarrow} + f^{\dag}_{2\uparrow}f^{\phantom\dag}_{1\downarrow})\hat{\tau}^-
+ \tfrac{1}{\sqrt{2}}(f^{\dag}_{2\downarrow}f^{\phantom\dag}_{1\uparrow} + f^{\dag}_{1\downarrow}f^{\phantom\dag}_{2\uparrow})\hat{\tau}^+\Bigr] 
\end{split}
\end{equation}
where $\sigma = +/- \iff\; \uparrow/\downarrow$, and $\hat{H}_{\mathrm{eff}}$ is given by eq. \ref{heff}.

Fig. \ref{fig:Bfield_entropycomparison} shows results for  $S_\mathrm{imp}(T)$,
in which the parameters are chosen to correspond to the 2AIM with $U/\Gamma = 20$, as considered earlier in 
fig.\ref{fig:B0_entropyComparison}.
First, for comparison, the short-dashed line reproduces the 2IKM result from 
fig. \ref{fig:Bfield_varJ2_entropy} 
for $J_1 = J_2$, 
where the 2CK critical FP is stable and hence $S_\mathrm{imp}(T=0)=\tfrac{1}{2}\ln 2$.
Next, the dotted line in fig. 
\ref{fig:Bfield_entropycomparison} shows the behavior of the corresponding $H_3'$ model (direct inter-lead charge 
transfer) with $ V_{LR} = \sqrt{JJ_1J_2/U}$. Since $H_3'$ destabilizes the 2CK FP there is ultimately a 
crossover to the SC FP with vanishing residual entropy, but the crossover scale here is sufficiently small compared to $\tk^{\prime}$ that a $\half\ln 2$ entropy plateau remains visible. The temperature window over which the plateau exists can be optimised by tuning $h$ very slightly away from the critical $h_c$ of the corresponding 2IKM; this has already been performed for the dotted line in fig.\ref{fig:Bfield_entropycomparison}
(where $(h-h_c)/h_c \simeq 2\times 10^{-4}$).

Turning now to the 2AIM and $H_3$ models, we have repeated the process of searching for the widest $\tfrac{1}{2}\ln 2$ entropy plateau by varying $h$ around $h_c$. However our calculations 
show no 
such plateau for the parameters considered: the entropy always crosses directly from $\ln 2$ to zero, as illustrated by the solid and dashed lines in fig. \ref{fig:Bfield_entropycomparison}
(2AIM and $H_3$ models, respectively). This is not surprising, since it mirrors the $h=0$ behavior shown in 
fig. \ref{fig:B0_entropyComparison}: while the $H_3'$ model for $U/\Gamma = 20$ shows a distinct 2CK entropy plateau, the 2AIM and $H_3$ models do not.

\begin{figure}[t]
\begin{center}
\includegraphics[height=45mm]{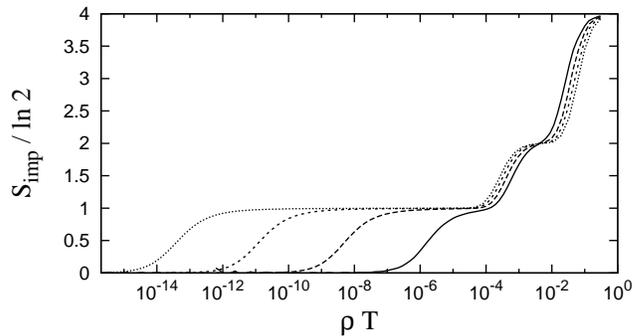}
\caption{$S_{\mathrm{imp}}(T)$ \emph{vs} $\rho T$ for the 2AIM with $h \simeq h_c$, $t/\Gamma = 1 \gg t_c/\Gamma$, and $U/\Gamma = 20$ (solid line), $30$ (long dash), $40$ (short dash) and $50$ (dotted). As $U/\Gamma$ increases, the $\ln 2$ entropy plateau extends to lower $T$, but for the values of $U/\Gamma$ considered no $\half\ln 2$  plateau is found.}
\label{fig:Bfield_varU_entropy}
\end{center}
\end{figure}

In sec.~\ref{CT} we explained that the 2CK FP can indeed be observed at $h=0$ in the 2AIM, if one considers the model at a larger $U/\Gamma \gtrsim 40$. It is thus natural to ask if the same is true for 2CK at finite field. We have undertaken preliminary calculations for $U/\Gamma$ up to $50$ (in practice the largest for which the calculations are feasible) in an effort to answer this question. In each case we have examined the entropy around $h\simeq h_c\sim J_{c}$ for signs of a $\tfrac{1}{2}\ln 2$ plateau but, as illustrated in 
fig. \ref{fig:Bfield_varU_entropy}, do not find any sign of 2CK behavior. This suggests that if 2CK is to be found in the 2AIM at finite field, it will arise only for $U/\Gamma$ in excess of 50, for which the corresponding $T_K'$ will surely be out of range of experimental grasp.


\section{Concluding remarks}

In this paper we have examined the possibility of observing two-channel Kondo physics in tunnel-coupled DQDs.
In the 
two-impurity Kondo model limit,
such physics clearly arises at zero magnetic field near the 2CK critical point, but in real quantum dots the effects of inter-lead charge transfer, which destabilise the 2CK fixed point, must of course be considered.

While direct inter-lead hopping generally destroys the 2CK physics on energy scales small 
compared to the Kondo scale,\cite{male:10} thus providing a seemingly large window over which 2CK behavior should be observable, we have argued that 
\emph{cotunneling} charge transfer processes -- proceeding \emph{via} the dot spins, and arising naturally within the 2AIM -- 
significantly reduce the likelihood of realising the 2CK physics experimentally.

A finite magnetic field opens up the possibility of a field-induced 2CK effect. 
For channel (LR-) symmetric systems, we showed
that the quantum phase transition of the zero-field 2IKM is the terminal point of a line of QPTs at finite field, the effective low energy critical model at large $h$ being a spin-anisotropic 2CK model.
Again, however, the charge transfer processes present in real TCDQDs turn the line of QPTs into a crossover; in this case their effect is even more destructive, and we find no evidence of field-induced 2CK physics in the 2AIM on experimentally-realisable energy scales. 

It has been proposed that longer, even-numbered quantum dot chains might be good candidate systems for observing 2CK physics.\cite{zara:06} Increasing the number of dots between the leads suppresses charge transfer, \cite{zara:06} 
but it also leads to a decrease\cite{mitc:09} 
in $\tk$, so that while the longer dot chain systems are more likely to flow to the 2CK FP ($\tk \gg T^*$), they 
are likely to do so at lower temperatures than for the two-dot case considered here. More work is needed to 
determine whether longer dot chains indeed offer a more promising route to accessing 2CK physics.

\begin{acknowledgments}
We are grateful to Mark Buitelaar for stimulating discussions, and to the EPSRC (U.K.) for financial support.
\end{acknowledgments}

%
\bibliography{paper}

\begin{thebibliography}{10}
\bibitem{note:phsymm}
This is largely for convenience. For the 2AIM away from ph symmetry, but with
  each dot in essence singly-occupied, potential scattering arises in the
  effective low-energy model. This does not destabilize the quantum phase
  transition in the 2IKM,\cite{jone:91} even with different potential
  scattering in each lead.\cite{zara:06} We do not therefore expect breaking of
  ph symmetry to affect our general conclusions.
\bibitem{note:alpha}
Otherwise there is a weak dependence of $\alpha$ on $\rho J_{1}$ itself.
\bibitem{note:Tstar}
In Ref. \onlinecite{male:10}, $T^*$ is calculated from the NRG energy levels.
  We have confirmed that our calculation of $T^*$ from $S_{\mathrm{imp}}$ gives
  essentially the same results.
\bibitem{note:cantran}
We also employ a trivial canonical transformation,\\ $c_{1\mathbf{k}\uparrow}
  \rightarrow -c_{1\mathbf{k}\uparrow}$ for $\uparrow$-spin electrons on lead
  $1$.
\end{thebibliography}
\end{document}